# DNA nanotechnology-enabled chiral plasmonics: from static to dynamic

*Chao Zhou[†], Xiaoyang Duan[†,‡], and Na Liu[†,‡]\**

[†] Max Planck Institute for Intelligent Systems, Heisenbergstrasse 3, D-70569 Stuttgart, Germany
[‡] Kirchhoff Institute for Physics, University of Heidelberg, Im Neuenheimer Feld 227, D-69120, Heidelberg, Germany

## Conspectus

The development of DNA nanotechnology, especially the advent of DNA origami, makes DNA ideally suited to construct nanostructures with unprecedented complexity and arbitrariness. As a fully addressable platform, DNA origami can be used to organize discrete entities in space through DNA hybridization with nanometer accuracy. Among a variety of functionalized particles, metal nanoparticles such as gold nanoparticles (AuNPs) feature an important pathway to endow DNA origami-assembled nanostructures with tailored optical functionalities. When metal particles are placed in close proximity, their particle plasmons, *i.e.*, collective oscillations of conduction electrons, can be coupled together, giving rise to a wealth of interesting optical phenomena. Nevertheless, characterization methods that are capable to read out the optical responses from plasmonic nanostructures composed of small metal particles, and especially to *in situ* optically distinguish their minute conformation changes, are very few. Circular dichroism (CD) spectroscopy has proven to be a successful means to overcome the challenges due to its high sensitivity on discrimination of three-dimensional conformation changes.

In this Account, we discuss a variety of static and dynamic chiral plasmonic nanostructures enabled by DNA nanotechnology. In the category of static plasmonic systems, we first show chiral plasmonic nanostructures based on spherical AuNPs, including plasmonic helices, toroids, and tetramers. To enhance the CD responses, anisotropic gold nanorods with larger extinction coefficients are utilized to create chiral plasmonic crosses and helical superstructures. Next, we highlight the inevitable evolution from static to dynamic plasmonic systems along with the fast development of this interdisciplinary field. Several dynamic plasmonic systems are reviewed according to their working mechanisms. We first elucidate a reconfigurable plasmonic cross structure, which can execute DNA-regulated conformational changes on the nanoscale. Hosted by a reconfigurable DNA origami, the plasmonic cross can be switched between a chiral locked state and an achiral relaxed state through toehold-mediated strand displacement reactions. This reconfigurable nanostructure can also be modified in response to light stimuli, leading to a noninvasive, waste-free, and all-optically controlled system. Taking one step further, we show that selective manipulations of individual structural species coexisting in one ensemble can be achieved using pH tuning of reconfigurable plasmonic nanostructures in a programmable manner. Finally, we describe an alternative to achieving dynamic plasmonic systems by driving AuNPs directly on origami. Such plasmonic walkers, inspired by the biological molecular motors in living cells, can generate dynamic CD responses when carrying out directional, progressive, and reverse nanoscale walking on DNA origami. We envision that the combination of DNA nanotechnology and plasmonics opens an avenue towards a new generation of functional plasmonic systems with tailored optical properties and useful applications including polarization conversion devices, biomolecular sensing, surface-enhanced Raman and fluorescence spectroscopy as well as diffraction-limited optics.

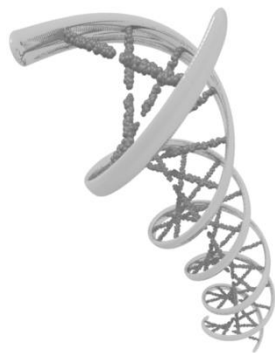 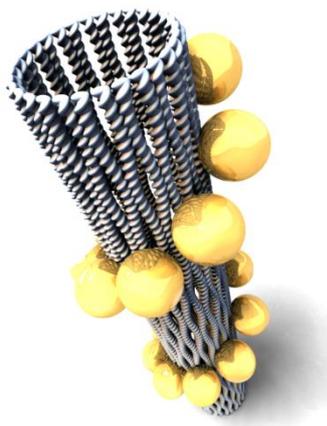

natural chirality    plasmonic chirality

# 1 Introduction

In 1980s, Nadrian Seeman brought DNA as construction material into the nanoscale world.[1] Several decades later in 2006, a crucial breakthrough in the field of DNA nanotechnology came with the concept called DNA origami, which was developed by Paul Rothemund.[2] The formation of DNA origami involves the folding of a long scaffold DNA strand by hundreds of short staple strands into nanostructures with nearly arbitrary shapes.[3-6] As each staple strand possesses a unique sequence and its position in the formed structure is deterministic, DNA origami is a fully addressable platform, which can be used to spatially organize discrete entities through DNA hybridization with nanometer accuracy.[7]

Among a variety of functionalized particles that can be assembled on DNA origami, metal nanoparticles such as silver or gold nanoparticles (AuNPs) feature an important pathway to endow DNA origami-assembled nanostructures with tailored optical functionalities. When light interacts with a metal nanoparticle, collective oscillations of conduction electrons known as particle plasmons are excited.[8] The resulting plasmon resonance critically depends on the material composition, the dielectric environment, the shape, and the size of the nanoparticle. When they are placed in close proximity, the possibilities for shaping and controlling near-field and far-field optical properties increase enormously.[8-10] Near-field coupling between metal nanoparticles is extremely sensitive to the conformation changes of the constructed plasmonic nanostructure. Such strong dependence on conformation provides great opportunities in manipulating optical responses on the nanoscale.

The DNA origami technique offers unique opportunities to create plasmonic architectures with complex two-dimensional (2D) and three-dimensional (3D) geometries.[11-23] Particularly, this technique enables the realization of dynamic plasmonic nanostructures, which can exhibit immediate conformational changes and thus dynamic optical response upon regulated external inputs.[24-30] This has remained challenging for other state of the art nanofabrication techniques. Nevertheless, characterization methods that are capable to read out the optical responses and especially to *in situ* optically distinguish the minute conformation changes of plasmonic nanostructures assembled using DNA origami are very few. The difficulties lie in threefold. First, the metal nanoparticles that can be assembled on DNA origami with a high yield are quite small, typically in the range of 5-30 nm in diameter. In the quasi-static regime, the scattering and absorption cross-sections are proportional to the six and third powers of the particle diameter, respectively.[31] Such plasmonic assemblies composed of small metal particles lead to weak optical responses with low signal to noise ratios, especially from scattering measurements. Second, the DNA origami technique allows for large-scale and even bulk production of plasmonic nanostructures in a highly parallel manner. Often, the as-fabricated plasmonic nanostructures are characterized in an aqueous solution. In other words, the assemblies are oriented randomly in an ensemble. This sets an inevitable constraint for the available optical characterization approaches, *i.e.*, the measured data should not be smeared out resulting from the averaged optical signals along all possible orientations. Third, the utilized optical approaches should be greatly sensitive on conformations, particularly for characterization of dynamic plasmonic nanostructures, so that the optical signal changes can be readily correlated with dynamic structural changes in real time. In practice, circular dichroism (CD) spectroscopy fulfills all the aforementioned requirements and has been successfully applied to optically characterize both static and dynamic DNA-assembled plasmonic nanostructures. In turn, DNA-assembled chiral plasmonic nanostructures are not only interesting as model systems for exploring profound physics on chiral interactions[32-34] but also may find useful applications, especially in chiral sensing[35,36].

In this Account, we will first introduce the general concept of plasmonic chirality. Subsequently, we will discuss a variety of chiral plasmonic nanostructures assembled using DNA origami. We will start with static systems and then transit to the newly developed dynamic systems. We will end this Account with conclusion and outlook.

## 2 Plasmonic chirality

Chiral objects are ubiquitous in nature. A large number of natural molecules that are essential to our lives are chiral, for example, amino-acids, carbohydrates, nucleic acids, proteins, *etc*. As our own hands are chiral themselves (see Fig. 1A), a chiral object is also termed "handed". The word chirality is actually derived from the Greek χειρ (*kheir*), "hand". Generally speaking, chirality is a symmetry property of an object, which does not possess mirror planes or inversion symmetry. The object and its mirror image called enantiomers cannot coincide by simple rotations or translations.

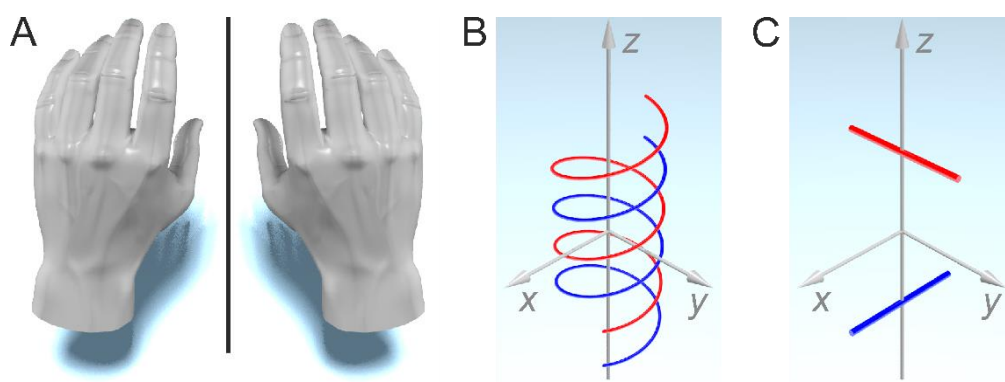

**Figure 1**. Schematic illustrations of representative chiral objects and chiral model systems. (A) Human hands; (B) Helix model system developed by Kauzmann; (C) Noncollinear cross model system developed by Born and Kuhn.

In optical spectroscopy, CD measurements are often employed to detect subtle conformation changes of chiral molecules.[37,38] This is due to the fact that molecular enantiomers interact differently with left-handed (LH) and right-handed (RH) circularly polarized light in absorption. Such an absorption difference, namely, CD, is a valuable means to distinguish the chiral states of molecules in modern pharmaceutics and drug industry. The microscopic origin of this optical activity lies in the nonlocality of the light-matter interaction. Two main classical approaches have been proposed to model the optical activity of chiral molecules. One is the helix model developed by Kauzmann (see Fig. 1B),[39] in which one electron is bound on a helix, giving rise to electric and magnetic characters to the optical transitions. It has been successfully applied to describe the chiral properties of helical polymers. The other is the coupled oscillator model developed by Born and Kuhn (see Fig. 1C).[40,41] The chiral response in this case results from the coupling of two electric dipoles arranged in a handed fashion, which is known to be the main mechanism in chiral molecules comprising two or more interacting monomers.

However, molecular CD of natural chiral molecules is typically very weak and occurs in the UV region (150–300 nm). Theoretical calculations pioneered by A. Govorov have predicted that plasmonic assemblies consisting of metal nanoparticles arranged in chiral geometries, for example, helix and noncollinear cross, can lead to strong optical chirality in the visible range.[42,43] At resonance, the dipolar plasmons of individual metal nanoparticles can be strongly coupled, resulting in collective plasmons that oscillate within the entire chiral architecture. Such plasmonic chirality can be artificially engineered not only in strength but also in wavelength position and handedness.[44,45]

Recently, different DNA-based approaches have been directed toward organizing metal nanoparticles into 3D chiral geometries.[46-51] For example, one strategy is associated with self-assembly of DNA tubules through integration of AuNPs.[47] Stacked rings, spirals, and nested spiral tubes were obtained utilizing size-dependent steric repulsion effects among AuNPs. Nevertheless, chiral structures of a certain handedness were not readily separated from the product mixture. In 2009, Mastroianni et al. demonstrated a chiral grouping of four different-sized AuNPs, which were monofunctionalized with distinct strands of DNA.[46] Due to the substantial particle size difference and structural non-rigidity, plasmonic chirality was not anticipated. In contrast, the DNA origami technique offers a compelling alternative to constructing chiral plasmonic nanostructures with strong chiral responses, high structural fidelity as well as remarkable structural complexity and rigidity. In the past several years, the area of DNA origami-assembled chiral plasmonic nanostructures has been extremely active and fruitful, bringing about many fascinating prototype systems with great potentials for applications. It is noteworthy that in the following, we will focus on plasmonic chirality resulting from near-field coupling among AuNPs. Plasmon-induced chirality that occurs due to Coulombic interactions between DNA and AuNPs will be neglected[52], given its much weaker response when compared to that of plasmonic chirality.

## 3 Static chiral plasmonic systems

The helix geometry might be the most intuitive blueprint to implement chiral plasmonic nanostructures. In collaboration with Ding's group, we created plasmonic helices composed of spherical AuNPs (10 nm) on tubular DNA origami (Fig. 2A).[15] The AuNPs functionalized with DNA strands were first positioned through DNA hybridization along two liner chains on rectangular DNA origami, followed by rationally rolling with the help of DNA folding strands. The AuNPs were then organized into a 3D helical geometry. Meanwhile, Kuzyk and co-workers introduced a more robust approach to create plasmonic helices, in which nine spherical AuNPs (10 nm) were arranged on a DNA origami bundle in a staircase fashion (see Fig. 2B).[14] The LH plasmonic helices exhibited a characteristic peak-dip CD profile in the visible wavelength range, while the RH helices showed a mirrored CD spectrum, agreeing well with the theoretical calculations.

In view of the averaged CD signals in solution, which hamper chiral plasmonic nanostructures for on-chip applications, we also demonstrated plasmonic toroids that exhibited distinct chiroptical properties along the axial orientation by simple spin-coating of the structures on a substrate as shown in Fig. 2C.[20] The nanostructure was assembled by linking four curved plasmonic helices. In total, 24 AuNPs were arranged along a DNA origami ring following a toroidal geometry. Given the unique circular symmetry, birefringence effects could be avoided.

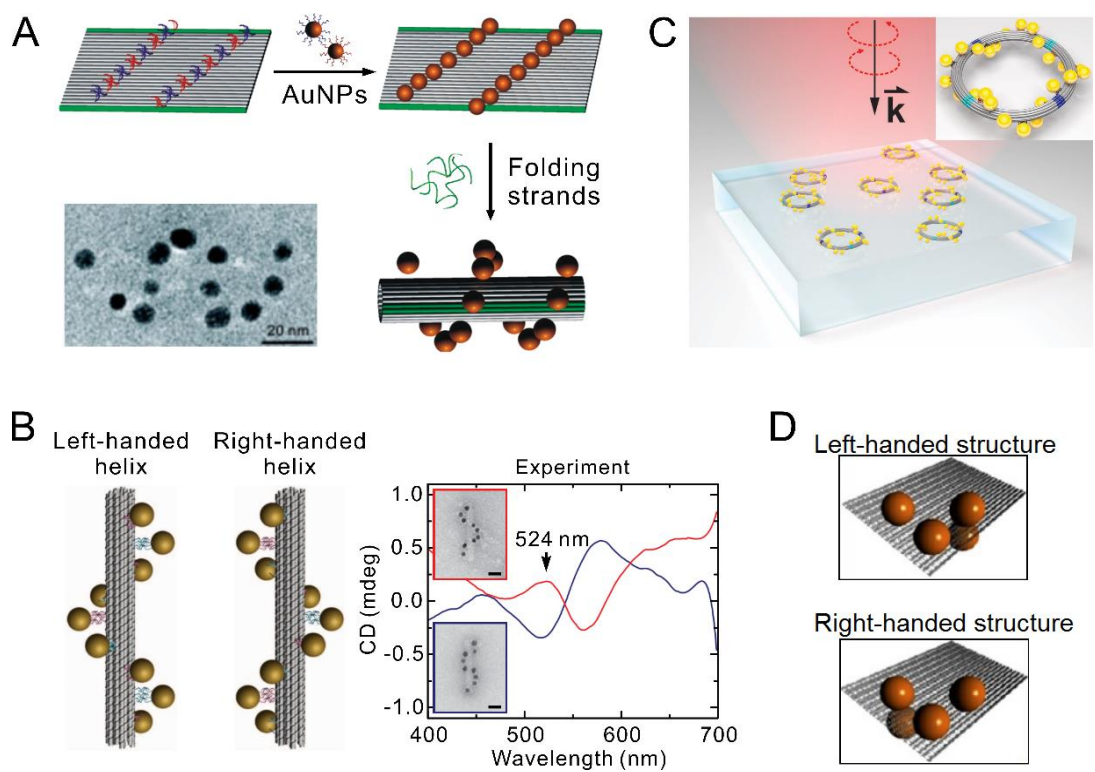

**Figure 2.** DNA origami-assembled chiral plasmonic nanostructures composed of spherical AuNPs. (A) Plasmonic helices formed by rolling a AuNP decorated origami sheet into an origami tube (reproduced with permission from ref. 15. Copyright 2011 American Chemical Society); (B) Plasmonic helices created by arranging AuNPs on origami bundles and the measured CD spectra (reproduced with permission from ref. 14.

Copyright 2012 Nature Publishing Group); (C) Plasmonic toroidal nanostructures dispersed on a glass substrate (reproduced with permission from ref. 20. Copyright 2016 American Chemical Society); (D) Plasmonic tetramers assembled on rectangular DNA origami in LH and RH, respectively (reproduced with permission from ref. 17. Copyright 2013 American Chemical Society).

The tetrahedron geometry constitutes another inspiration for constructing chiral plasmonic nanostructures. In stereochemistry, a tetrahedral carbon atom that has four different substituents can form a chiral center. Simple organization of four different-sized AuNPs in tetrahedron geometry, however, would not lead to substantial CD signals due to weak coupling as discussed in the previous session. Alternatively, we assembled four identical AuNPs in an asymmetric tetrahedron to enable collective oscillations of plasmons within the structure.[17] The AuNPs were rationally arranged on a 2D origami sheet. Three of them were positioned along an L-line on one side of the origami (see Fig. 2D). The fourth AuNP was positioned on the other side, directly below one of the three AuNPs. Depending on the position of the fourth AuNP, the LH or RH plasmonic enantiomer could be achieved.

The dominating role of spherical AuNPs as key player for building DNA-assembled chiral plasmonic nanostructures was terminated after the successful demonstration of well-controlled gold nanorod (AuNR) assemblies on origami by Pal *et al*.[12] Due to their stronger optical strength and anisotropic nature, AuNRs are excellent candidates for the realization of advanced plasmonic architectures with distinct and tailored optical functionalities.[53] Our group and Wang's group both demonstrated plasmonic cross nanostructures.[16,18] A single-layer DNA origami sheet was employed to assemble two AuNRs (40 nm × 12 nm) in different planes, forming a 90º twisting angle (see Fig. 3A). The two crossed AuNRs constituted a 3D plasmonic chiral object, which generated a theme of handedness when interacting with LH and RH circularly polarized light, giving rise to strong CD (see Fig. 3A). With a comparable concentration, the CD intensity at resonance was over ten times stronger than that from chiral plasmonic nanostructures composed of spherical AuNPs. Even the CD feature at a shorter wavelength resulting from the transverse mode of the AuNRs was experimentally observable (see the black framed region). Taking a step further, Wang's group realized plasmonic helical superstructures as shown in Fig. 3B.[19] Capture strands in an 'X' pattern were arranged on the two sides of a more rigid double-layer origami template. AuNRs functionalized with complementary DNA strands were positioned on the origami and subsequently led to AuNR helices with the origami intercalated between neighboring AuNRs. LH and RH AuNR helices were conveniently accomplished by tuning the mirrored-symmetric 'X' patterns of the capture strands on the origami.

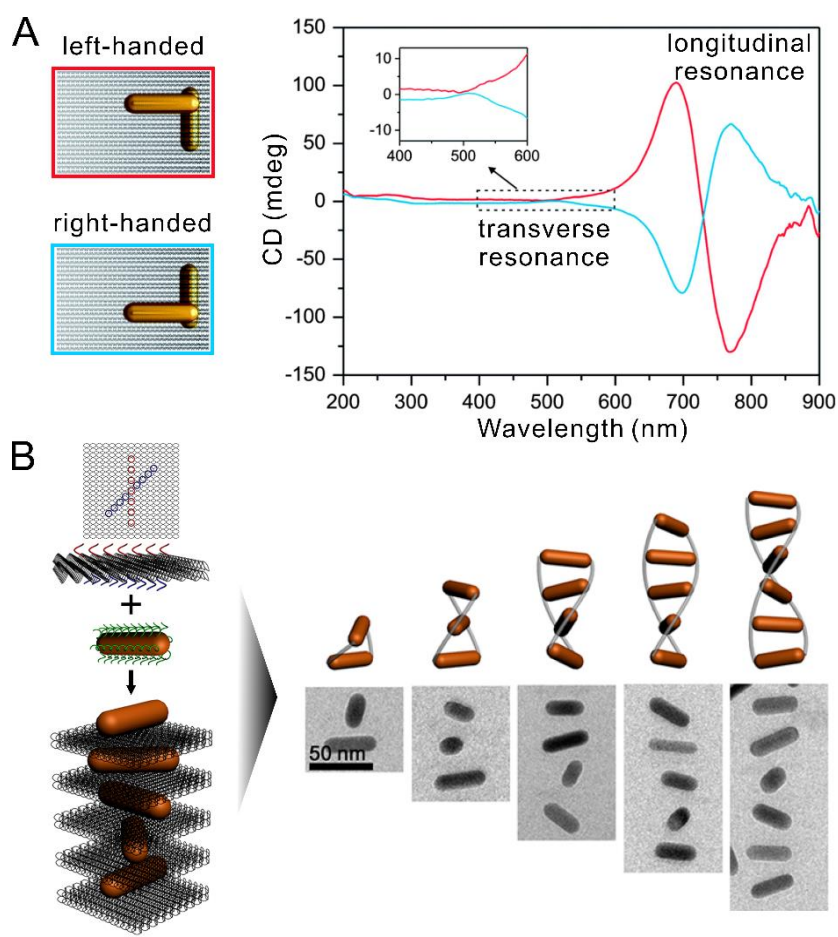

**Figure 3.** DNA origami-assembled chiral plasmonic nanostructures composed of anisotropic AuNRs. (A) Plasmonic cross nanostructures (reproduced with permission from ref. 18. Copyright 2014 The Royal Society of Chemistry); (B) Plasmonic helical superstructures (reproduced with permission from ref. 19. Copyright 2014 American Chemical Society).

# 4 Dynamic chiral plasmonic systems

Dynamic plasmonic devices that can be controlled by regulated physical or chemical inputs hold great promise for applications in adaptable nanophotonic circuitry and optical molecular sensing.[54-57] The DNA origami technique offers an unprecedented pathway to impart reconfigurability to passive systems in that DNA structures allow for various ways to control their dynamic behavior. Basic schemes for structural reconfigurations of DNA structures include toehold-mediated strand displacement reactions[58] as well as pH,[59] ion concentration,[60] magnetic, light,[61] and thermal stimuli[60] as illustrated in Fig. 4.

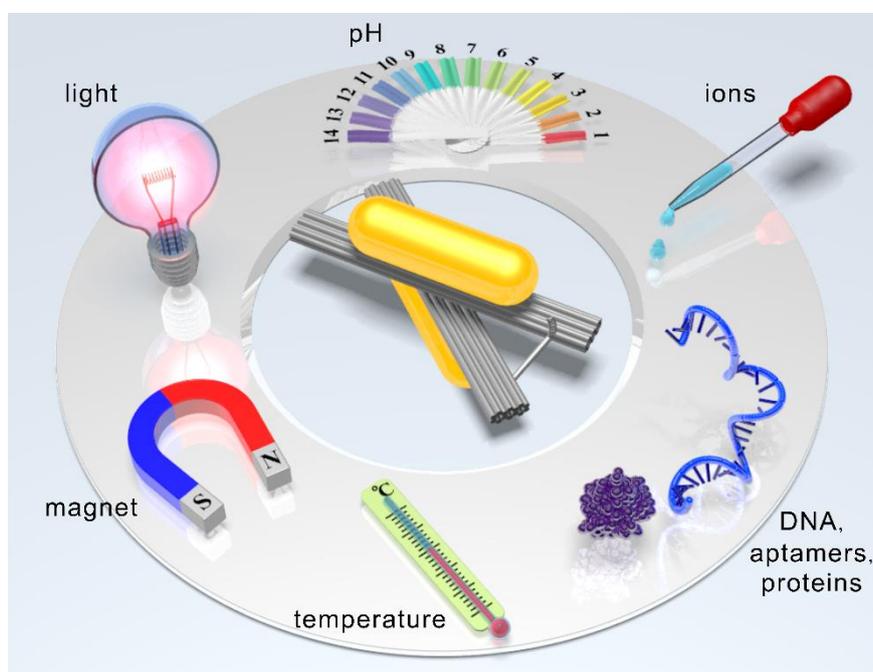

**Figure 4**. Schematic of possible means for driving DNA-based dynamic plasmonic systems.

## 4.1 DNA-fueled dynamic plasmonic nanostructures

Our group demonstrated one of the first reconfigurable plasmonic nanosystems, which could execute DNA-regulated conformational changes on the nanoscale.[24] Two AuNRs (40 nm × 10 nm) were hosted on a reconfigurable DNA origami template through hybridization (see Fig. 5A). Two DNA locks were incorporated in the origami with different strand sequences to enable independent controls (see Fig. 5B). In a relaxed state, both of the DNA locks were open. Arm *a* possessed a double-stranded DNA (dsDNA) segment (green and violet) with a single-stranded DNA (ssDNA) toehold (orange). Arm *b* possessed a dsDNA segment (grey) with a locking sequence (green). When removal strand 1 ($R_1$) was added, dissociation of the DNA strand with the toehold from arm *a* was triggered. Subsequently, the two DNA origami bundles were joined together through DNA hybridization and the system was directed to a LH state. Upon addition of return strand 1 ($\bar{R}_1$), the DNA lock was pulled open and the system returned to its relaxed state. Following a similar route, the system could be reconfigured between its relaxed and RH state upon addition of fuel strands ($R_2$ and $\bar{R}_2$). The switching processes were revealed by *in situ* monitoring the CD intensity at a fixed wavelength of 725 nm. As shown in Fig. 5C, switching among three different states could be realized in multiple cycles by successive additions of corresponding fuel strands.

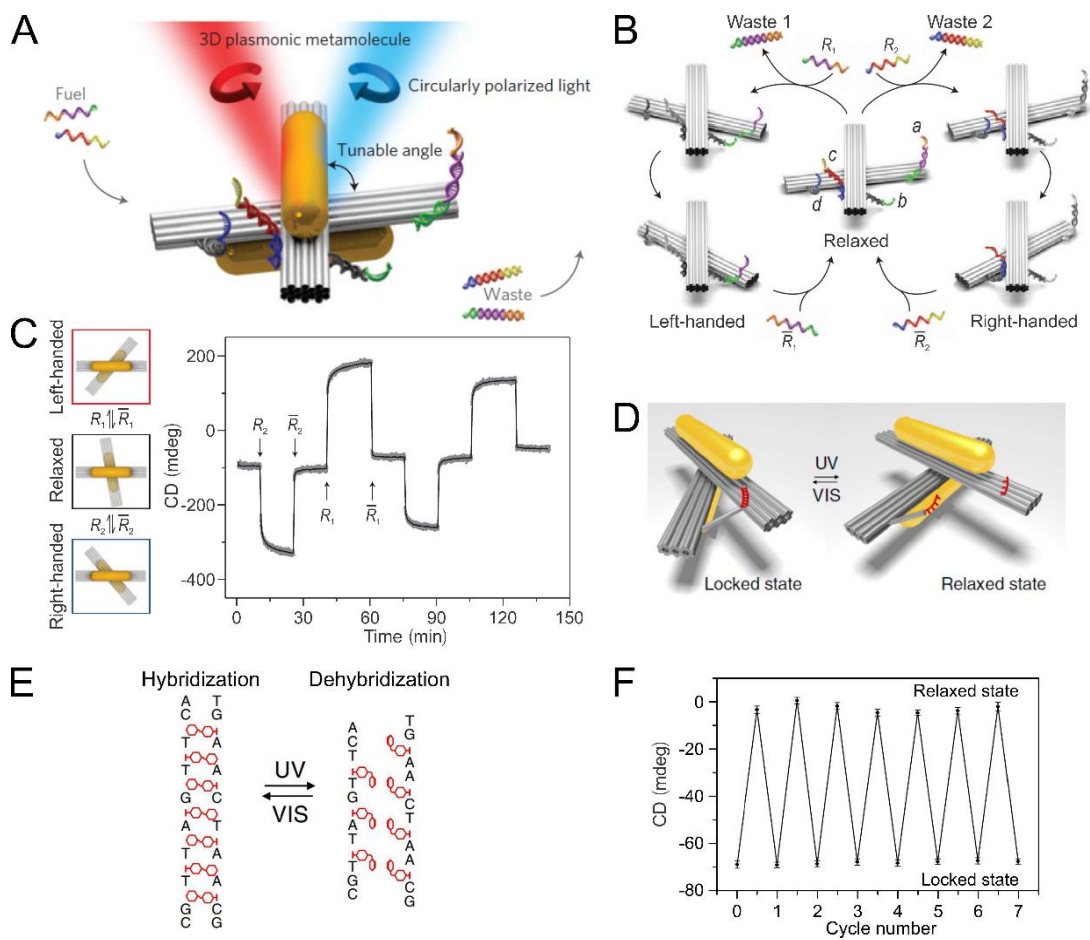

**Figure 5.** Reconfigurable chiral plasmonic systems. (A) Schematic of the reconfigurable plasmonic cross structure; (B) Switching mechanism involving the toehold-mediated strand displacement reactions; (C) *In situ* monitoring of the CD signals over time at a fixed wavelength of 725 nm during switching; (D) Schematic of the photoresponsive plasmonic cross nanostructure regulated by light for switching between the locked RH and relaxed states (reproduced with permission from ref. 27. Copyright 2016 Nature Publishing Group); (E) Light switching of the azobenzene-modified DNA lock; (F) CD switching in multiple cycles recorded at a fixed wavelength during alternative ultraviolet and visible light illumination. (Panel A-C reproduced with permission from ref. 24. Copyright 2014 Nature Publishing Group; Panel D-F reproduced with permission from ref. 27. Copyright 2016 Nature Publishing Group).

## 4.2 Light-driven dynamic plasmonic nanostructures

The strand displacement reactions generate DNA duplex waste in each cycle. Additions of fuel strands also cause intervention as well as dilution effects. In contrast, light as an energy input is noninvasive and waste free. Also, different from chemical fuels that crucially depend on diffusion kinetics, light offers high spatial and temporal resolution as it can be switched on and off rapidly. We designed an all-optically controlled dynamic plasmonic system by integration of a photoresponsive DNA lock on DNA origami (see Fig. 5D).[27] Azobenzene, a molecule that is well-known for its photoisomerization, could be incorporated into DNA strands for reversible control of DNA hybridization (see Fig. 5E).[61,62] As illustrated in Fig. 5D, each arm of the DNA lock contained a segment of azobenzene-modified oligonucleotides. The two segments were pseudocomplementary, meaning they could be hybridized to form a duplex when the azobenzene molecules were in the *trans*-form. In this case, the plasmonic nanostructure was locked in the RH state. Upon ultraviolet light illumination, the azobenzene molecules were transformed to the *cis*-form, resulting in the opening of the DNA lock. Consequently, the plasmonic nanostructure went to the relaxed state. Upon visible light illumination, the azobenzene molecules were converted to the *trans*-form and the system was again locked to the RH state. To monitor the dynamic process associated with the conformation changes triggered by light, the CD response of a plasmonic sample was measured after visible (450 nm) and UV (365 nm) light illumination, respectively (see Fig. 5F). Excellent switching between the two states was achieved, demonstrating reliable photo-responsivity of the dynamic plasmonic system.

## 4.3 pH-controlled dynamic plasmonic nanostructures

Very recently, we have demonstrated selective tuning of individual structural species coexisting within one ensemble using pH controls in a programmable manner (see Fig. 6A).[28] As shown in Fig. 6B, the pH-sensitive lock contained a DNA triplex formed through pH-dependent Hoogsteen interactions (dots) between ssDNA and a DNA duplex formed through Watson-Crick interactions (dashed). The transition pH value of a DNA lock between the locked and opened states could be precisely programmed over a wide range by simply varying the relative content of TAT/CGC triplets.[63] By integrating these DNA locks on reconfigurable DNA origami, we created a series of chiral plasmonic nanostructures that could be tuned to respond at different pH values (see Fig. 6C). These plasmonic nanostructures showed considerably narrow transition pH windows and rapid responses on the order of tens of seconds.

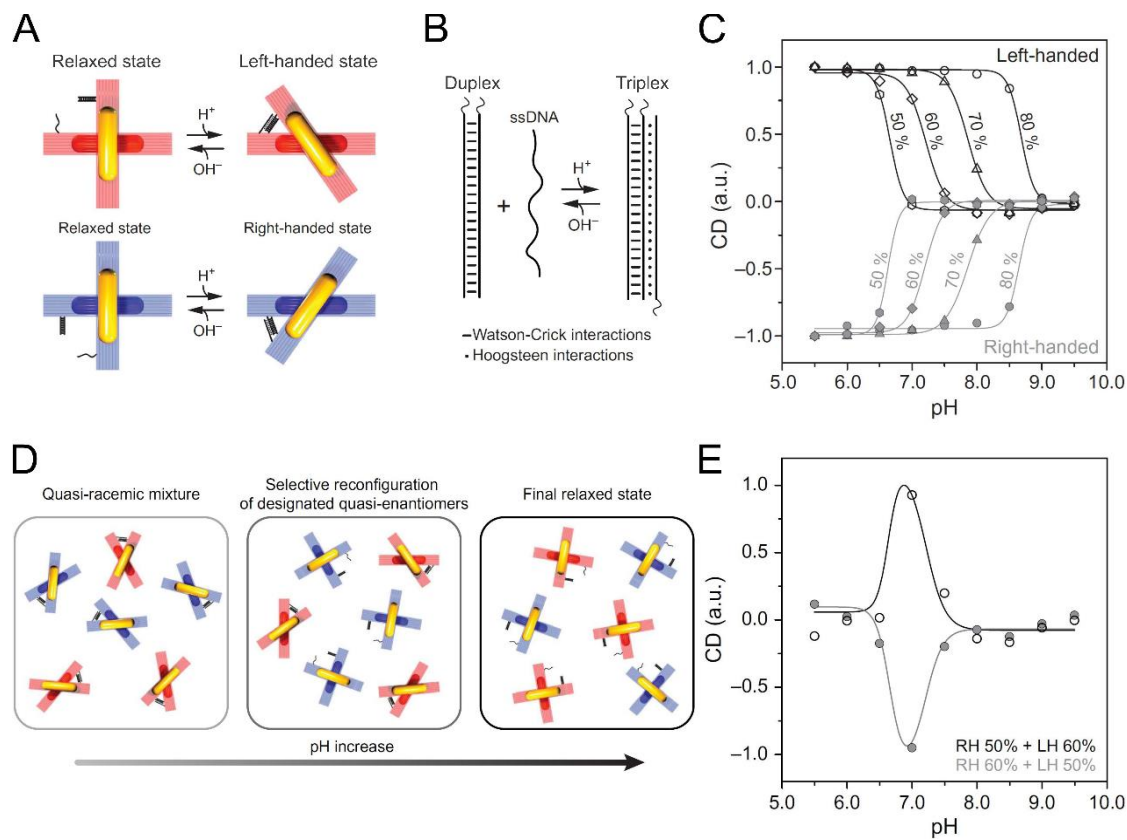

**Figure 6**. Reconfigurable chiral plasmonic nanostructures driven by pH changes. (A) Switching between the relaxed and the LH/RH states by opening or closing the pH-triggered DNA lock; (B) Schematic of the pH-triggered DNA lock; (C) Relative CD dependence on pH tuning for the LH/RH plasmonic nanostructures with DNA locks containing different TAT/CGC contents. Solid lines are the fitting results using Hill function; (D) Schematic of the selective reconfiguration of the designated quasi-enantiomers; (E) Relative CD dependence on pH tuning for a mixture of two quasi-enantiomers: RH 50% and LH 60% (black open circles), RH 60% and LH 50% (grey solid circles). (Reproduced with permission from ref. 28. Copyright 2017 AAAS).

The ability to engineer structural dynamics with programmable pH tuning opens a pathway towards discriminative control of different structural species mixed in a solution. As a proof of concept experiment, LH and RH plasmonic quasi-enantiomers containing pH-sensitive DNA locks with different relative contents of TAT/CGC triplets were mixed and selectively controlled by tuning the pH values over a wide range (see Fig. 6D). For example, RH nanostructures with DNA locks containing 50% TAT (transition pH~6.6) and LH ones containing 60% TAT (transition pH~7.2) were mixed in equimolar amounts. The CD intensity of this quasi-racemic mixture was nearly zero at pHs lower than 6.5 (see Fig. 6E), as both quasi-enantiomers were in the locked state. When the pH increased, the RH structures in the solution were gradually opened, going to the relaxed state. In contrast, LH 60% remained unresponsive in the solution. As a result, an abrupt CD peak was observed at pH 7.0. When the pH continued to increase, LH 60% became also responsive and the LH structures in the solution were opened gradually. The overall CD response of the mixture therefore decreased until pH 8. Similarly, the quasi-racemic solution containing LH 50% and RH 60% exhibited nearly mirrored pH-dependence.

## 4.4 Plasmonic walkers

In living cells, molecular motors such as kinesin and dynein can walk directionally along microtubules to ferry cargoes.[64] Such biological machines were the inspiration for the development of artificial DNA walkers[65-69] and plasmonic walkers[25,26], which could execute directional, progressive, and reverse nanoscale walking on DNA origami.[25] The plasmonic walker system consisted of a double-layer DNA origami track, a AuNR (yellow) as the walker and another AuNR (red) as the stator as shown in Fig. 7A. The walker AuNR was fully functionalized with identical DNA strands as feet. Six parallel rows of footholds (represented by A-F) were arranged on the track. In each row, there were five identical footholds as binding sites. All footholds contained the same segment (black part) for binding with the foot strands but footholds in different rows were distinguished by the toehold sections (colored parts). Five stations (I to V) were defined on the track with spacing of 7 nm, which was also the step size of the walker. Initially, the walker resided on foothold rows A and B at station I and the rest of the rows were deactivated by blocking strands.

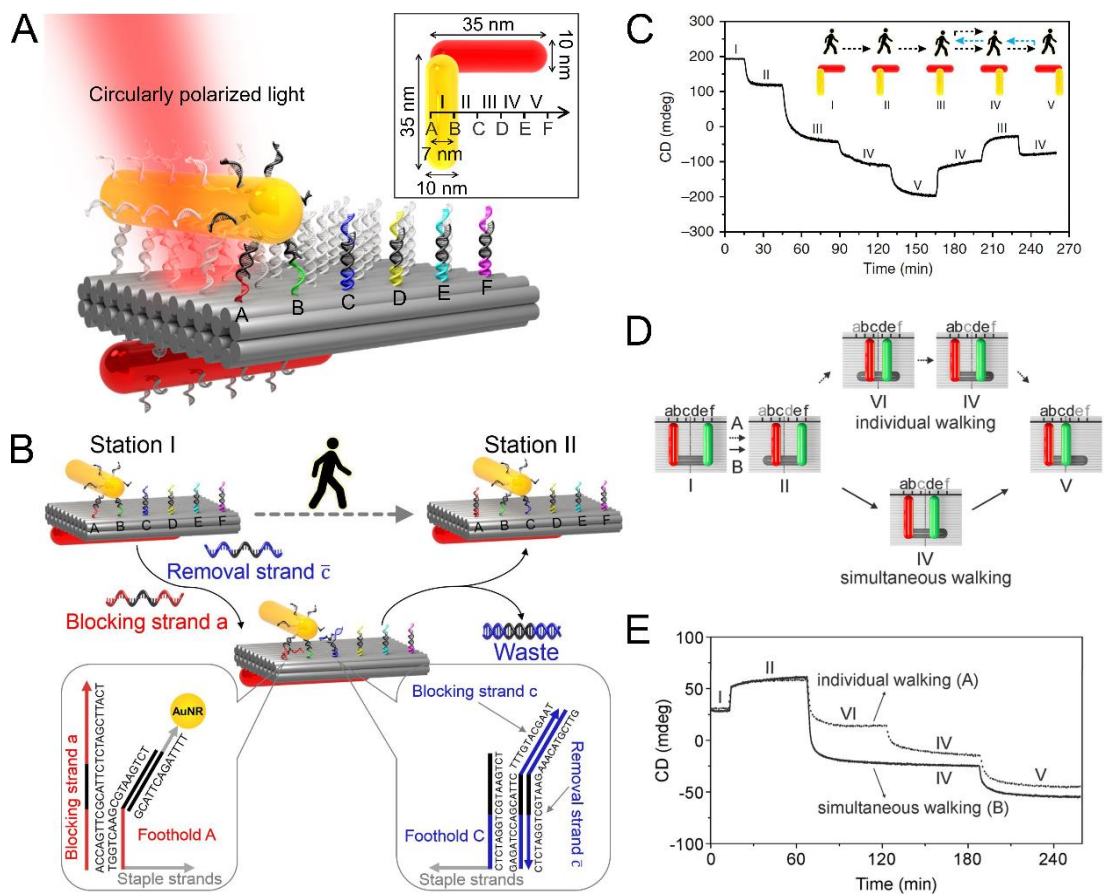

**Figure 7.** Chiral plasmonic walkers. (A) Schematic of the plasmonic walker; (B) Walking mechanism illustration; (C) CD intensity changes as the walker performs stepwise walking following a route I–II–III–IV–V–IV–III–IV; (D) Individual and simultaneous walking routes of the plasmonic walker couple; (E) Dynamic processes of individual and simultaneous walking monitored by *in situ* CD intensity changes. (Panel A-C reproduced with permission from ref. 25. Copyright 2015 Nature Publishing Group; Panel D and E reproduced with permission from ref. 26. Copyright 2015 American Chemical Society).

After one step walking, the walker arrived at station II, stepping on foothold rows B and C. Such walking involved two processes, both of which were driven by toehold-mediated strand-displacement reactions (see Fig. 7B). By adding appropriate DNA fuels, the walker could impose progressive walking by alternatively advancing its feet in a 'rolling' manner. The discrete walking steps could be correlated with a series of CD signal changes, while the plasmonic walker was dynamically coupled with the stator during walking as shown in Fig. 7C. Two AuNR walkers could be also accommodated on the same track, giving rise to a plasmonic walker couple system as shown in Fig. 7D. The CD responses were collectively determined by the positions of both walkers relative to the stator.[26] Individual walking of each walker as well as simultaneous walking of both walkers could be optically distinguished by CD spectroscopy (see Fig. 7E).

## 5 Conclusion and outlook

The combination of knowhow in DNA nanotechnology and plasmonics opens an avenue towards a new generation of functional plasmonic devices. Thanks to the versatile design possibilities, full addressability, and sub-10 nm resolution, DNA origami offers an exceptional platform to create plasmonic nanostructures with well-defined geometries and interesting optical properties. Due to the fact that DNA structures can respond to strand-displacement reactions, DNA aptamer-target interactions, DNA-protein interactions and so forth, dynamic plasmonic nanostructures are promising candidates for many biological sensing applications with remarkable target specificity and modularity. For example, a single bio-entity that binds to the DNA lock of a chiral plasmonic system and its subsequent reactions with other chemical compounds could trigger a series of conformational changes, which can be detected by single structure CD spectroscopy in real time. Single structure spectroscopy can eliminate inhomogeneous broadening that ensemble measurements often encounter, therefore offering an effective way to *in situ* detect individual dynamic behavior and processes. In addition, reconfigurable 3D plasmonic chiral nanostructures may lead to advanced optical metamaterials, whose refractive indices could be altered by a variety of external means between positive and negative values.[70,71] It will also be of great interest, if DNA-assembled plasmonic structures can be functionalized on substrates with ordered orientations but still retain structural reconfigurability so that metasurfaces[72] that possess dynamic phase distributions can be achieved to exhibit a variety of novel functionalities.

Meanwhile, there is plenty of room for explorations to further advance this exciting multidisciplinary field forward. Bottleneck problems still remain and are waiting for solutions. For instance, how to enhance the stability of DNA-assembled plasmonic structures over time? How to functionalize large plasmonic particles on origami with high fidelity and high yield for stronger optical responses? How to scale up origami sizes to accommodate more elements for achieving complex systems? When plasmonic particles are placed in extremely close proximity, *i.e.*, on the order of several nanometers or less, does one expect quantum effects and how to interpret them? Does DNA play any essential role in such optical processes? These are very relevant open questions that certainly deserve a lot of brilliant ideas and experimental efforts.

## AUTHOR INFORMATION

**Corresponding Author**

*E-mail: laura.liu@is.mpg.de

**Author Contributions**

The manuscript was written by all authors. All authors have given approval to the final version of the manuscript.

**Notes**

The authors declare no competing financial interest.

**Biographical Information**

**Chao Zhou** joined Prof. Liu's group as a postdoc at the Max Planck Institute for Intelligent Systems, Germany in 2013.

**Xiaoyang Duan** is a Ph. D student in Prof. Liu's group at University of Heidelberg, Germany.

**Na Liu** is Professor at the Kirchhoff Institute for Physics at University of Heidelberg, Germany.

ACKNOWLEDGMENT

We acknowledge support from the Sofja Kovalevskaja grant from the Alexander von Humboldt-Foundation, the Volkswagen grant, and the European Research Council (ERC Dynamic Nano) grant.

REFERENCES

1. Seeman, N. C. Nucleic-Acid Junctions and Lattices. *J. Theor. Biol.* **1982**, *99*, 237-247.
2. Rothemund, P. W. K. Folding DNA to Create Nanoscale Shapes and Patterns. *Nature* **2006**, *440*, 297-302.
3. Dietz, H.; Douglas, S. M.; Shih, W. M. Folding DNA into Twisted and Curved Nanoscale Shapes. *Science* **2009**, *325*, 725-730.
4. Douglas, S. M.; Dietz, H.; Liedl, T.; Hogberg, B.; Graf, F.; Shih, W. M. Self-Assembly of DNA into Nanoscale Three-Dimensional Shapes. *Nature* **2009**, *459*, 414-418.
5. Han, D. R.; Pal, S.; Nangreave, J.; Deng, Z. T.; Liu, Y.; Yan, H. DNA Origami with Complex Curvatures in Three-Dimensional Space. *Science* **2011**, *332*, 342-346.
6. Wang, P. F.; Meyer, T. A.; Pan, V.; Dutta, P. K.; Ke, Y. G. The Beauty and Utility of DNA Origami. *Chem* **2017**, *2*, 359-382.
7. Hong, F.; Zhang, F.; Liu, Y.; Yan, Y. DNA Origami: Scaffolds for Creating Higher Order Structures. *Chem. Rev.* **2017**. DOI: http://dx.doi.org/ 10.1021/acs.chemrev.6b00825.
8. Halas, N. J.; Lal, S.; Chang, W. S.; Link, S.; Nordlander, P. Plasmons in Strongly Coupled Metallic Nanostructures. *Chem. Rev.* **2011**, *111*, 3913-3961.
9. Lal, S.; Link, S.; Halas, N. J. Nano-Optics from Sensing to Waveguiding. *Nat. Photon.* **2007**, *1*, 641-648.
10. Sonnichsen, C.; Reinhard, B. M.; Liphardt, J.; Alivisatos, A. P. A Molecular Ruler Based on Plasmon Coupling of Single Gold and Silver Nanoparticles. *Nat. Biotechnol.* **2005**, *23*, 741-745.
11. Ding, B. Q.; Deng, Z. T.; Yan, H.; Cabrini, S.; Zuckermann, R. N.; Bokor, J. Gold Nanoparticle Self-Similar Chain Structure Organized by DNA Origami. *J. Am. Chem. Soc.* **2010**, *132*, 3248-3249.
12. Pal, S.; Deng, Z. T.; Wang, H. N.; Zou, S. L.; Liu, Y.; Yan, H. DNA Directed Self-Assembly of Anisotropic Plasmonic Nanostructures. *J. Am. Chem. Soc.* **2011**, *133*, 17606-17609.
13. Tan, S. J.; Campolongo, M. J.; Luo, D.; Cheng, W. L. Building Plasmonic Nanostructures with DNA. *Nat. Nanotech.* **2011**, *6*, 268-276.
14. Kuzyk, A.; Schreiber, R.; Fan, Z. Y.; Pardatscher, G.; Roller, E. M.; Hogele, A.; Simmel, F. C.; Govorov, A. O.; Liedl, T. DNA-Based Self-Assembly of Chiral Plasmonic Nanostructures with Tailored Optical Response. *Nature* **2012**, *483*, 311-314.
15. Shen, X. B.; Song, C.; Wang, J. Y.; Shi, D. W.; Wang, Z. A.; Liu, N.; Ding, B. Q. Rolling up Gold Nanoparticle-Dressed DNA Origami into Three-Dimensional Plasmonic Chiral Nanostructures. *J. Am. Chem. Soc.* **2012**, *134*, 146-149.
16. Lan, X.; Chen, Z.; Dai, G.; Lu, X.; Ni, W.; Wang, Q. Bifacial DNA Origami-Directed Discrete, Three-Dimensional, Anisotropic Plasmonic Nanoarchitectures with Tailored Optical Chirality. *J. Am. Chem. Soc.* **2013**, *135*, 11441-11444.

17. Shen, X. B.; Asenjo-Garcia, A.; Liu, Q.; Jiang, Q.; de Abajo, F. J. G.; Liu, N.; Ding, B. Q. Three-Dimensional Plasmonic Chiral Tetramers Assembled by DNA Origami. *Nano Lett.* **2013**, *13*, 2128-2133.
18. Shen, X. B.; Zhan, P. F.; Kuzyk, A.; Liu, Q.; Asenjo-Garcia, A.; Zhang, H.; de Abajo, F. J. G.; Govorov, A.; Ding, B. Q.; Liu, N. 3D Plasmonic Chiral Colloids. *Nanoscale* **2014**, *6*, 2077-2081.
19. Lan, X.; Lu, X. X.; Shen, C. Q.; Ke, Y. G.; Ni, W. H.; Wang, Q. B. Au Nanorod Helical Superstructures with Designed Chirality. *J. Am. Chem. Soc.* **2015**, *137*, 457-462.
20. Urban, M. J.; Dutta, P. K.; Wang, P. F.; Duan, X. Y.; Shen, X. B.; Ding, B. Q.; Ke, Y. G.; Liu, N. Plasmonic Toroidal Metamolecules Assembled by DNA Origami. *J. Am. Chem. Soc.* **2016**, *138*, 5495-5498.
21. Shen, C. Q.; Lan, X.; Zhu, C. G.; Zhang, W.; Wang, L. Y.; Wang, Q. B. Spiral Patterning of Au Nanoparticles on Au Nanorod Surface to Form Chiral AuNR@AuNP Helical Superstructures Templated by DNA Origami. *Adv. Mater.* **2017**, *29*. DOI: http://dx.doi.org/10.1002/adma.201606533.
22. Tian, Y.; Wang, T.; Liu, W. Y.; Xin, H. L.; Li, H. L.; Ke, Y. G.; Shih, W. M.; Gang, O. Prescribed Nanoparticle Cluster Architectures and Low-Dimensional Arrays Built Using Octahedral DNA Origami Frames. *Nat. Nanotech.* **2015**, *10*, 637-644.
23. Cecconello, A.; Kahn, J. S.; Lu, C. H.; Khorashad, L. K.; Govorov, A. O.; Willner, I. DNA Scaffolds for the Dictated Assembly of Left-/Right-Handed Plasmonic Au Np Helices with Programmed Chiro-Optical Properties. *J. Am. Chem. Soc.* **2016**, *138*, 9895-9901.
24. Kuzyk, A.; Schreiber, R.; Zhang, H.; Govorov, A. O.; Liedl, T.; Liu, N. Reconfigurable 3D Plasmonic Metamolecules. *Nat. Mater.* **2014**, *13*, 862-866.
25. Zhou, C.; Duan, X.; Liu, N. A Plasmonic Nanorod That Walks on DNA Origami. *Nat. Commun.* **2015**, *6*, 8102.
26. Urban, M. J.; Zhou, C.; Duan, X. Y.; Liu, N. Optically Resolving the Dynamic Walking of a Plasmonic Walker Couple. *Nano Lett.* **2015**, *15*, 8392-8396.
27. Kuzyk, A.; Yang, Y. Y.; Duan, X.; Stoll, S.; Govorov, A. O.; Sugiyama, H.; Endo, M.; Liu, N. A Light-Driven Three-Dimensional Plasmonic Nanosystem That Translates Molecular Motion into Reversible Chiroptical Function. *Nat. Commun.* **2016**, *7*, 10591.
28. Kuzyk, A.; Urban, M. J.; Idili, A.; Ricci, F.; Liu, N. Selective Control of Reconfigurable Chiral Plasmonic Metamolecules. *Sci. Adv.* **2017**, *3*, e1602803.
29. Schreiber, R.; Luong, N.; Fan, Z. Y.; Kuzyk, A.; Nickels, P. C.; Zhang, T.; Smith, D. M.; Yurke, B.; Kuang, W.; Govorov, A. O.; Liedl, T. Chiral Plasmonic DNA Nanostructures with Switchable Circular Dichroism. *Nat. Commun.* **2013**, *4*, 2948.
30. Qian, Z. X.; Ginger, D. S. Reversibly Reconfigurable Colloidal Plasmonic Nanomaterials. *J. Am. Chem. Soc.* **2017**, *139*, 5266-5276.
31. Maier, S. A.: Plasmonics: Fundamentals and Applications; Springer US: New York, 2007.
32. Govorov, A. O.; Fan, Z. Y.; Hernandez, P.; Slocik, J. M.; Naik, R. R. Theory of Circular Dichroism of Nanomaterials Comprising Chiral Molecules and Nanocrystals: Plasmon Enhancement, Dipole Interactions, and Dielectric Effects. *Nano Lett.* **2010**, *10*, 1374-1382.
33. Fan, Z. Y.; Govorov, A. O. Helical Metal Nanoparticle Assemblies with Defects: Plasmonic Chirality and Circular Dichroism. *J. Phys. Chem. C* **2011**, *115*, 13254-13261.

34. Fan, Z. Y.; Zhang, H.; Govorov, A. O. Optical Properties of Chiral Plasmonic Tetramers: Circular Dichroism and Multipole Effects. *J. Phys. Chem. C* **2013**, *117*, 14770-14777.
35. Xu, Z.; Xu, L. G.; Liz-Marzan, L. M.; Ma, W.; Kotov, N. A.; Wang, L. B.; Kuang, H.; Xu, C. L. Sensitive Detection of Silver Ions Based on Chiroplasmonic Assemblies of Nanoparticles. *Adv Opt Mater* **2013**, *1*, 626-630.
36. Tang, L. J.; Li, S.; Xu, L. G.; Ma, W.; Kuang, H.; Wang, L. B.; Xu, C. L. Chirality-Based Au@Ag Nanorod Dimers Sensor for Ultrasensitive Psa Detection. *ACS Appl. Mater. Interfaces* **2015**, *7*, 12708-12712.
37. Fasman, G. D.: Circular Dichroism and the Conformational Analysis of Biomolecules; Springer US: New York, 1996.
38. Berova, N.; Nakanishi, K.; Woody, R. W.: Circular Dichroism: Principles and Applications, 2nd Edition; Wiley-VCH: New York, 2000.
39. Kauzmann, W.: Quantum Chemistry; Academic Press: New York, 1957.
40. Born, M. The Natural Optical Activity of Fuidity and Gases. *Phys. Z.* **1915**, *16*, 251-258.
41. Kuhn, W. Quantative Ratio and Connections in Natural Optical Activity. *Z. Phys. Chem.* **1929**, *4*, 14-36.
42. Fan, Z. Y.; Govorov, A. O. Plasmonic Circular Dichroism of Chiral Metal Nanoparticle Assemblies. *Nano Lett.* **2010**, *10*, 2580-2587.
43. Yan, W. J.; Xu, L. G.; Xu, C. L.; Ma, W.; Kuang, H.; Wang, L. B.; Kotov, N. A. Self-Assembly of Chiral Nanoparticle Pyramids with Strong R/S Optical Activity. *J. Am. Chem. Soc.* **2012**, *134*, 15114-15121.
44. Ma, W.; Xu, L. G.; de Moura, A. F.; Wu, X. L.; Kuang, H.; Xu, C. L.; Kotov, N. A. Chiral Inorganic Nanostructures. *Chem. Rev.* **2017**, *117*, 8041-8093.
45. Hentschel, M.; Schaferling, M.; Duan, X. Y.; Giessen, H.; Liu, N. Chiral Plasmonics. *Sci. Adv.* **2017**, *3*, e1602735.
46. Mastroianni, A. J.; Claridge, S. A.; Alivisatos, A. P. Pyramidal and Chiral Groupings of Gold Nanocrystals Assembled Using DNA Scaffolds. *J. Am. Chem. Soc.* **2009**, *131*, 8455-8459.
47. Sharma, J.; Chhabra, R.; Cheng, A.; Brownell, J.; Liu, Y.; Yan, H. Control of Self-Assembly of DNA Tubules through Integration of Gold Nanoparticles. *Science* **2009**, *323*, 112-116.
48. Li, Z. T.; Zhu, Z. N.; Liu, W. J.; Zhou, Y. L.; Han, B.; Gao, Y.; Tang, Z. Y. Reversible Plasmonic Circular Dichroism of Au Nanorod and DNA Assemblies. *J. Am. Chem. Soc.* **2012**, *134*, 3322-3325.
49. Ma, W.; Kuang, H.; Wang, L. B.; Xu, L. G.; Chang, W. S.; Zhang, H. N.; Sun, M. Z.; Zhu, Y. Y.; Zhao, Y.; Liu, L. Q.; Xu, C. L.; Link, S.; Kotov, N. A. Chiral Plasmonics of Self-Assembled Nanorod Dimers. *Sci. Rep.* **2013**, *3*, 1934.
50. Cecconello, A.; Besteiro, L. V.; Govorov, A. O.; Willner, I. Chiroplasmonic DNA-Based Nanostructures. *Nat. Rev. Mater.* **2017**, *2*, 17039.
51. Liu, H.; Shen, X. B.; Wang, Z. G.; Kuzyk, A.; Ding, B. Q. Helical Nanostructures Based on DNA Self-Assembly. *Nanoscale* **2014**, *6*, 9331-9338.
52. Govorov, A. O. Plasmon-Induced Circular Dichroism of a Chiral Molecule in the Vicinity of Metal Nanocrystals. Application to Various Geometries. *J. Phys. Chem. C* **2011**, *115*, 7914-7923.
53. Auguie, B.; Alonso-Gomez, J. L.; Guerrero-Martinez, A.; Liz-Marzan, L. M. Fingers Crossed: Optical Activity of a Chiral Dimer of Plasmonic Nanorods. *J. Phys. Chem. Lett.* **2011**, *2*, 846-851.

54. MacDonald, K. F.; Samson, Z. L.; Stockman, M. I.; Zheludev, N. I. Ultrafast Active Plasmonics. *Nat. Photon.* **2009**, *3*, 55-58.
55. Zheng, Y. B.; Yang, Y. W.; Jensen, L.; Fang, L.; Juluri, B. K.; Flood, A. H.; Weiss, P. S.; Stoddart, J. F.; Huang, T. J. Active Molecular Plasmonics: Controlling Plasmon Resonances with Molecular Switches. *Nano Lett.* **2009**, *9*, 819-25.
56. Zhang, S.; Zhou, J. F.; Park, Y. S.; Rho, J.; Singh, R.; Nam, S.; Azad, A. K.; Chen, H. T.; Yin, X. B.; Taylor, A. J.; Zhang, X. Photoinduced Handedness Switching in Terahertz Chiral Metamolecules. *Nat. Commun.* **2012**, *3*, 942.
57. Sun, Y. H.; Jiang, L.; Zhong, L. B.; Jiang, Y. Y.; Chen, X. D. Towards Active Plasmonic Response Devices. *Nano Res.* **2015**, *8*, 406-417.
58. Zhang, D. Y.; Seelig, G. Dynamic DNA Nanotechnology Using Strand-Displacement Reactions. *Nat. Chem.* **2011**, *3*, 103-113.
59. Dong, Y. C.; Yang, Z. Q.; Liu, D. S. DNA Nanotechnology Based on I-Motif Structures. *Acc. Chem. Res.* **2014**, *47*, 1853-1860.
60. Gerling, T.; Wagenbauer, K. F.; Neuner, A. M.; Dietz, H. Dynamic DNA Devices and Assemblies Formed by Shape-Complementary, Non-Base Pairing 3D Components. *Science* **2015**, *347*, 1446-1452.
61. Kamiya, Y.; Asanuma, H. Light-Driven DNA Nanomachine with a Photoresponsive Molecular Engine. *Acc. Chem. Res.* **2014**, *47*, 1663-1672.
62. Asanuma, H.; Liang, X.; Nishioka, H.; Matsunaga, D.; Liu, M.; Komiyama, M. Synthesis of Azobenzene-Tethered DNA for Reversible Photo-Regulation of DNA Functions: Hybridization and Transcription. *Nat. Protoc.* **2007**, *2*, 203-212.
63. Idili, A.; Vallee-Belisle, A.; Ricci, F. Programmable Ph-Triggered DNA Nanoswitches. *J. Am. Chem. Soc.* **2014**, *136*, 5836-5839.
64. Vale, R. D. The Molecular Motor Toolbox for Intracellular Transport. *Cell* **2003**, *112*, 467-480.
65. Sherman, W. B.; Seeman, N. C. A Precisely Controlled DNA Biped Walking Device. *Nano Lett.* **2004**, *4*, 1801-1801.
66. Shin, J. S.; Pierce, N. A. A Synthetic DNA Walker for Molecular Transport. *J. Am. Chem. Soc.* **2004**, *126*, 10834-10835.
67. Omabegho, T.; Sha, R.; Seeman, N. C. A Bipedal DNA Brownian Motor with Coordinated Legs. *Science* **2009**, *324*, 67-71.
68. Gu, H. Z.; Chao, J.; Xiao, S. J.; Seeman, N. C. A Proximity-Based Programmable DNA Nanoscale Assembly Line. *Nature* **2010**, *465*, 202-205.
69. Lund, K.; Manzo, A. J.; Dabby, N.; Michelotti, N.; Johnson-Buck, A.; Nangreave, J.; Taylor, S.; Pei, R. J.; Stojanovic, M. N.; Walter, N. G.; Winfree, E.; Yan, H. Molecular Robots Guided by Prescriptive Landscapes. *Nature* **2010**, *465*, 206-210.
70. Pendry, J. B. A Chiral Route to Negative Refraction. *Science* **2004**, *306*, 1353-1355.
71. Zhang, S.; Park, Y. S.; Li, J. S.; Lu, X. C.; Zhang, W. L.; Zhang, X. Negative Refractive Index in Chiral Metamaterials. *Phys. Rev. Lett.* **2009**, *102*.
72. Yu, N. F.; Capasso, F. Flat Optics with Designer Metasurfaces. *Nat. Mater.* **2014**, *13*, 139-150.